\documentclass[a4paper,12pt]{amsart}

\usepackage{kantlipsum} % for text filler
\calclayout{}

\usepackage[ruled,vlined]{algorithm2e}
\usepackage{amsfonts}
\usepackage{amsmath}    % need for subequations
\usepackage{amssymb}    %useful mathematical symbols
\usepackage{bm}         %needed for bold greek letters and math symbols
\usepackage{bbm}
\usepackage{color}
\usepackage{enumitem}
\usepackage{graphicx}   % need for PS figures
\usepackage{hyperref}
\usepackage{natbib}    %number and author-year style referencing
\usepackage{epsf}
\usepackage{lscape}
\usepackage{enumitem}
\usepackage{longtable}
\bibpunct{(}{)}{;}{a}{,}{,}

\begin{document}

\title{Discussion of {\it  Martingale Posterior Distributions}}
\author{David Rossell}

\maketitle

\date{}  %comment to include current date

%Max 400 words

Here is a brief discussion of  {\it Martingale posterior distributions}, by \cite{fong:2021}.
Congratulations on a thought-provoking piece. Building Bayesian inference from a (likelihood, predictive) pair, rather than a (likelihood, prior), enriches the paradigm and provides new ways to think about, formulate and solve problems. A few respectful remarks.

\begin{figure}
\begin{center}
\begin{tabular}{cc}
Bernoulli & Gaussian \\
\includegraphics[width=0.5\textwidth]{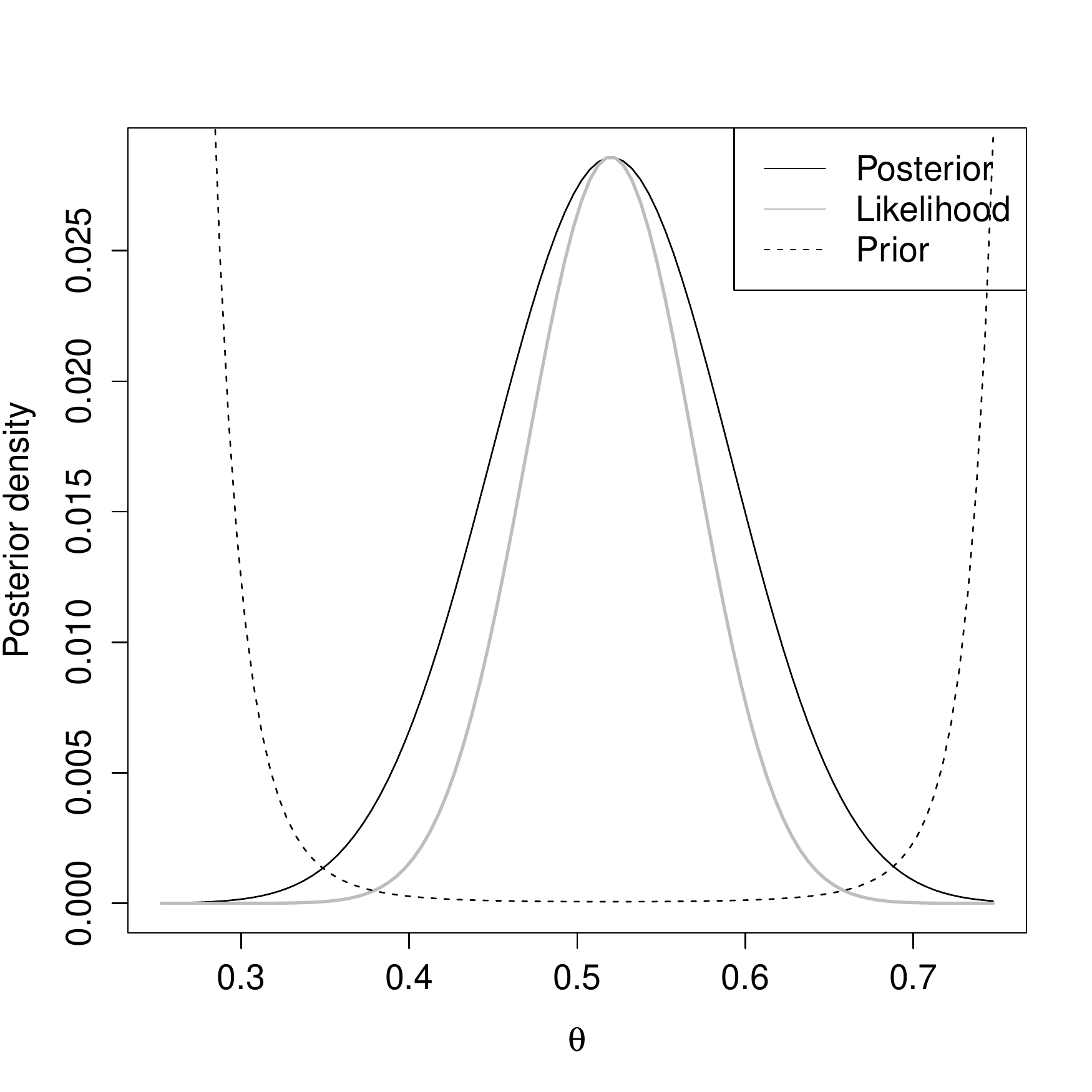} &
\includegraphics[width=0.5\textwidth]{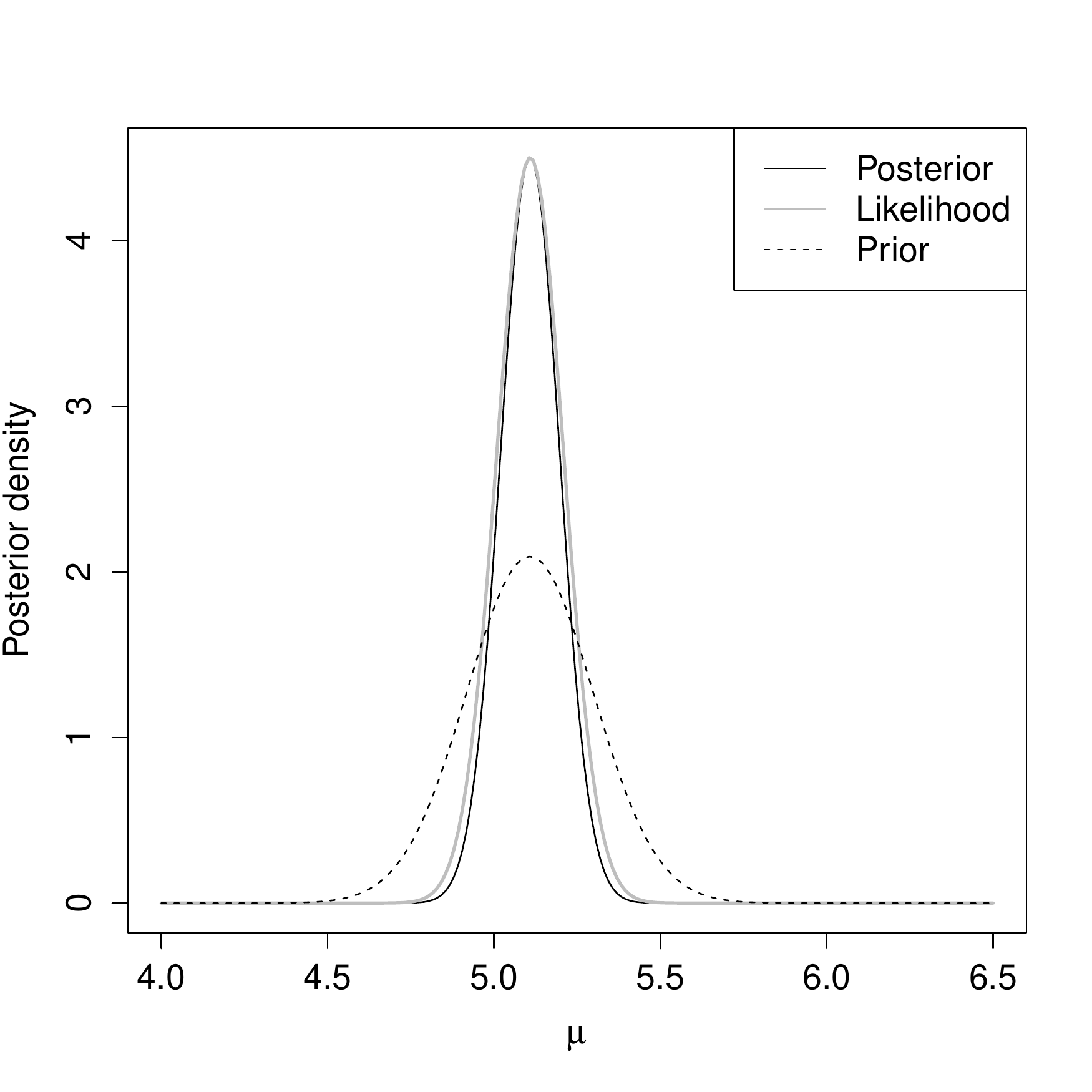} \\
\end{tabular}
\end{center}
\caption{Likelihood, posterior and prior densities for Bernoulli(0.5) and Normal(5,1) examples with $n=100$}
\label{fig:priorplot}
\end{figure}

First, although the authors never claim this, it is worth emphasizing that the framework is not prior-free. There is a posterior and a likelihood, hence the prior is proportional to their ratio.
The key is that said prior is data-dependent, providing an interesting avenue to develop objective Bayes methods, at the cost of loosing the coherence property in belief updating.
Inspecting the prior can be informative. 
Figure \ref{fig:priorplot} shows a Bernoulli example where truly $\theta=0.5$ but the implied prior places little mass around that value, and a Gaussian example where the prior is centered around the sample mean\footnote{code at \url{https://github.com/davidrusi/paper\_examples/tree/main/2022\_Rossell\_martingale\_posteriors})}.
This apparently erratic prior behavior might be problematic for model choice via Bayes factors, e.g. returning a very small integrated likelihood in the Bernoulli example.

Second, while sometimes it is easier to elicit a predictive than a prior, in my experience the reverse is often true. For example in regression a prior on parameters defines a prior on the $R^2$ coefficient, an easy-to-interpret quantity, whereas eliciting predictives may be less intuitive for non-statisticians. Further, note that computational considerations elegantly discussed by the authors severely restrict the range of predictives one may consider in practice, limiting the flexibility of the framework.

Third, I am afraid I disagree on the frameworks' computational convenience. Doing a single optimization may be faster than sampling, but the framework requires solving many optimizations. This is not cheaper than posterior sampling in a standard (likelihood, prior) construction, also the latter offers fast non-sampling based tools, e.g. Laplace approximations and extensions. It would be interesting to consider analogues for the predictive framework.

Finally, a remark on assuming that at  $n=\infty$ there is no uncertainty left. In some settings this is not true, e.g. in high-dimensional regression with $p \gg n$ (one adds higher-order polynomial terms as $n$ grows, say) and a normal prior on the parameters there remains posterior uncertainty even as $n \rightarrow \infty$. The proposed framework does not account for such uncertainty, unless suitable adjustments are made.

%\begin{figure}
%\begin{center}
%\begin{tabular}{cc}
%$n=10$ & $n=100$ \\
%\includegraphics[width=0.5\textwidth]{figs/bern_prior_n10_dataset1.pdf} &
%\includegraphics[width=0.5\textwidth]{figs/bern_prior_n100_dataset1.pdf} \\
%\includegraphics[width=0.5\textwidth]{figs/bern_prior_n10_dataset2.pdf} &
%\includegraphics[width=0.5\textwidth]{figs/bern_prior_n100_dataset2.pdf}
%\end{tabular}
%\end{center}
%\caption{Bernoulli simulation. Posterior (Dirichlet-bootstrap), likelihood and implied prior}
%\label{fig:bern_prior}
%\end{figure}
% 
% 
%\begin{figure}
%\begin{center}
%\begin{tabular}{cc}
%$n=10$ & $n=100$ \\
%\includegraphics[width=0.5\textwidth]{figs/prior_n10_dataset1.pdf} &
%\includegraphics[width=0.5\textwidth]{figs/prior_n100_dataset1.pdf} \\
%\includegraphics[width=0.5\textwidth]{figs/prior_n10_dataset2.pdf} &
%\includegraphics[width=0.5\textwidth]{figs/prior_n100_dataset2.pdf}
%\end{tabular}
%\end{center}
%\caption{Gaussian likelihood, posterior ($10^5$ Dirichlet-bootstrap draws with posterior predictive sample $N=10^4$) and implied prior}
%\label{fig:gaussian_prior}
%\end{figure}

\bibliographystyle{plainnat}
\bibliography{references}

\end{document}